\documentclass[
  twocolumn
]{scrartcl}
\usepackage[varg]{txfonts}   % Web of Conferences font
\usepackage{hyperref}
\usepackage{cleveref}
\usepackage{siunitx}
\usepackage{pgf}

\addtokomafont{title}{\huge}
\addtokomafont{author}{\large}
\addtokomafont{publishers}{\large}

\hypersetup{
  pdftitle={Pythia 8 as hadronic interaction model in air shower simulations},
  pdfauthor={Maximilian Reininghaus, Torbjörn Sjöstrand, Marius Utheim},
  pdflang=en
}

\usepackage[
  backend=biber,
  style=phys,
  biblabel=brackets,
  articletitle=false,
  doi=false,
  eprint=true,
  isbn=true,
  url=true,
  giveninits=true,
]{biblatex}
\newcommand{\Xmax}{X_{\mathrm{max}}}
\newcommand{\Nmu}{N_{\mu}}

\title{Pythia 8 as hadronic interaction model \\ in air shower simulations}
\author{Maximilian~Reininghaus\textsuperscript{1} for the CORSIKA~8~Collaboration\footnote{full author list available at
\url{https://s.kit.edu/c8-authorlist-uhecr2022}},\\
Torbjörn~Sjöstrand\textsuperscript{2},
Marius~Utheim\textsuperscript{3}}

\date{}

\publishers{
{\textsuperscript{1}\,Karlsruher Institut für Technologie (KIT), Postfach 3640, 76021 Karlsruhe, Germany}\\
{\textsuperscript{2}\,Department of Physics, Lund University, Sölvegatan 14A, S-223 62 Lund, Sweden}\\
{\textsuperscript{3}\,University of Jyvaskyla, Department of Physics, P.O. Box 35, FI-40014 University of Jyvaskyla, Finland}
}

\begin{document}
\maketitle
\begin{abstract}%
  \itshape
  Hadronic interaction models are a core ingredient of simulations of extensive air showers and pose the major source of uncertainties
  of predictions of air shower observables. Recently, Pythia~8, a hadronic interaction model popular in accelerator-based high-energy
  physics, became usable in air shower simulations as well. We have integrated Pythia~8 with its new capabilities into the air shower
  simulation framework CORSIKA~8. First results show significantly shallower shower development, which we attribute to higher cross-section
  predictions by the new simplified nuclear model of Pythia.
\end{abstract}
%\end{@twocolumnfalse}

\section{Introduction}
Large-scale experiments in modern ultra-high energy cosmic ray research rely heavily on simulations of extensive air showers
to link air shower observables to properties of the primary cosmic ray. An important ingredient in these simulations are
hadronic interaction models (also known as event generators) that govern the interactions of hadronic shower particles with air nuclei. Due to the nature
of the strong interaction, the wealth of hadronic interactions and multiparticle production is difficult or impossible to calculate from first principles alone. Only hard processes, i.e.\ those involving a large momentum transfer, can be treated within the framework of
perturbative quantum chromodynamics (pQCD). The bulk of soft interactions, however, relies mainly on phenomenological modelling
in combination with theoretical constraints and pQCD~\cite{Engel:2011zzb}. The most widely used up-to-date models used in air
shower simulations are EPOS-LHC~\cite{Pierog:2013ria}, QGSJet-II.04~\cite{Ostapchenko:2010vb} and SIBYLL~2.3d~\cite{Engel:2019dsg}.
While EPOS-LHC has its origins in heavy-ion physics, the other two are specifically tailored to the needs of air shower simulations
and the features of hadronic interactions important in that context.

In accelerator-based high-energy physics (HEP), a very popular event generator is Pythia~\cite{Sjostrand:2019zhc}, currently at
version 8.3~\cite{Bierlich:2022pfr}. For a long time, Pythia was not suitable for air shower simulations, mainly due to entirely
different setups employed in HEP and air shower simulations: While simulations for accelerators typically generate
a large number of events with the same settings (beam particle IDs and momenta), air shower simulations require event generation
with these settings randomly varying event by event. Recently, progress has been made in making Pythia~8 more suitable
for that setup~\cite{Sjostrand:2021dal}. On the one hand, this pertains to an accelerated context switching between beam
parameters. On the other hand, the number of valid beam particles has been extended and a simplified model of hadron-nucleus
collisions has been developed, while the fully-fledged Angantyr module~\cite{Bierlich:2018xfw} for heavy-ion collisions is not yet usable in air shower
simulations. Moreover, the range of valid beam energies has been extended down to $\sim \SI{200}{\MeV}$~(lab), which means
that Pythia can be used without an additional low-energy hadronic interaction model or just as such together with another
high-energy model.

In this contribution we describe and analyse first results obtained using Pythia~8 in a realistic air shower simulation. For this,
we have integrated Pythia~8.307 into the air shower simulation framework CORSIKA~8~\cite{Engel:2018akg} (Note that Pythia~8 already has been used to handle particle decays in CORSIKA~8). This work represents a
continuation of the study started in ref.~\cite{Reininghaus:2022gnr}.

\begin{figure*}[tb]
\centering
\begin{minipage}[b]{.45\textwidth}\centering
\includegraphics[width=\textwidth]{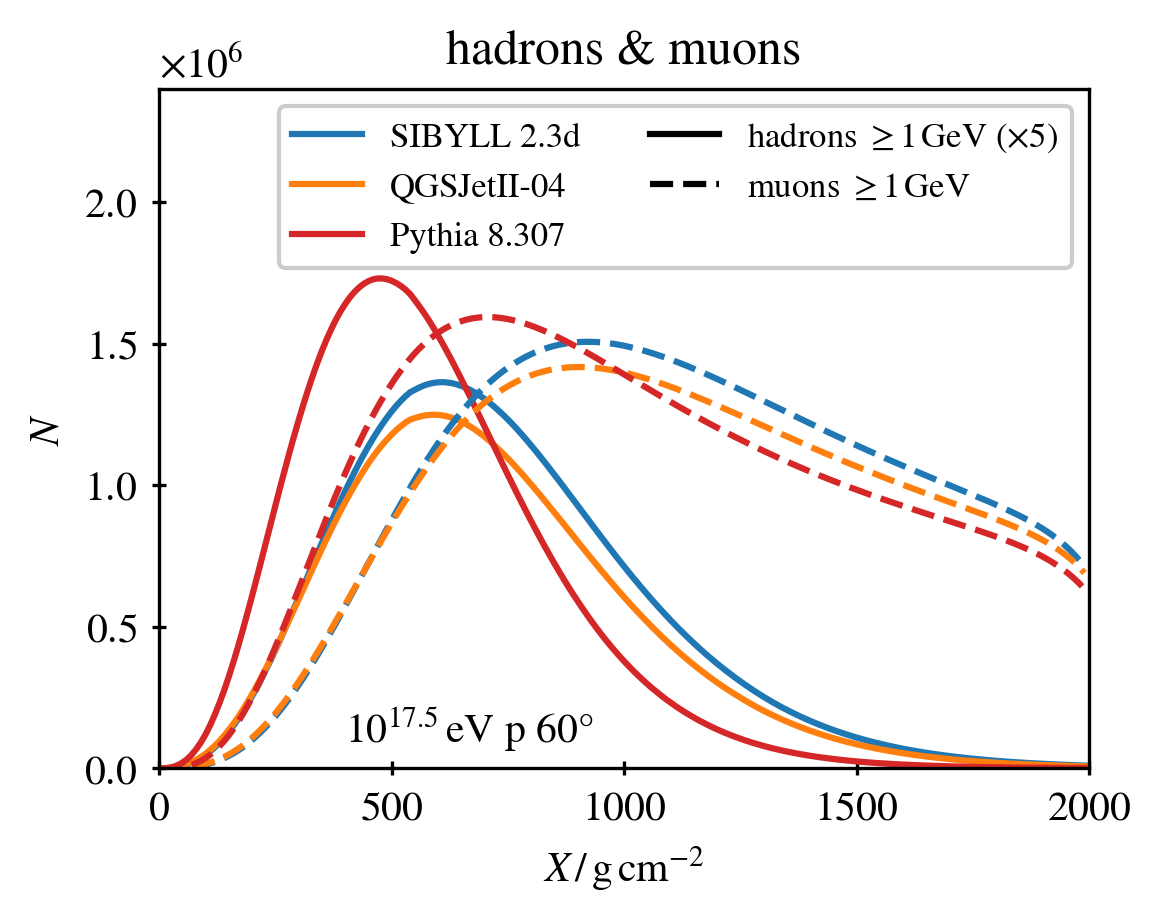}
\end{minipage}
\begin{minipage}[b]{.45\textwidth}\centering
\includegraphics[width=\textwidth]{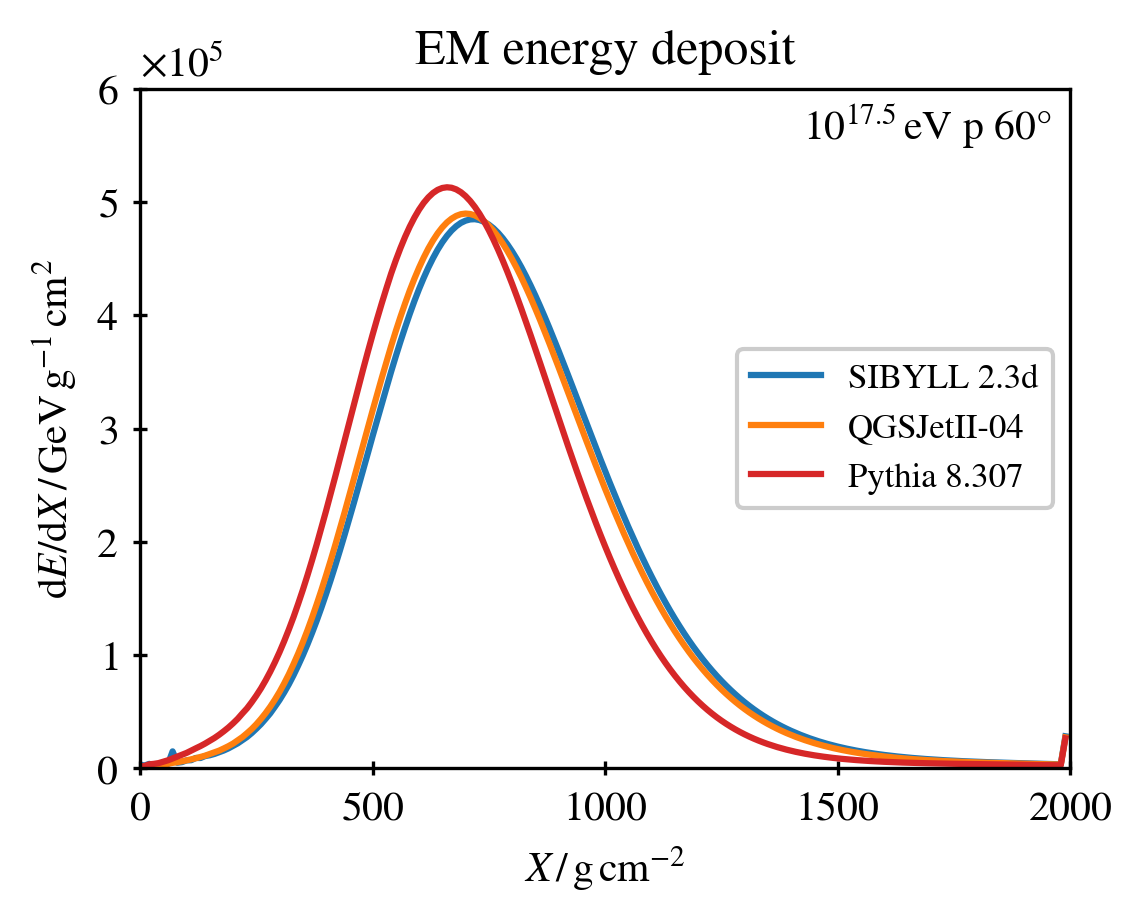}
\end{minipage}
\caption{Longitudinal profiles. Left: hadrons and muons. Right: electromagnetic energy deposit.}\label{fig:longprof}
\end{figure*}

\section{Setup}
We simulate air showers in a hybrid fashion: Hadronic interactions and propagation of hadron and muons are treated
in a Monte Carlo manner in CORSIKA~8. Electromagnetic particles are passed to CONEX~\cite{Bergmann:2006yz}, which generates
longitudinal profiles by solving the cascade equations numerically. Hadrons and muons stemming from photohadronic interactions
are therefore missing in this setup.
We consider showers at $10^{17.5}\,\si{\eV}$ with an
inclination of \ang{60} using Linsley's parameterization of the US Standard atmosphere~(see e.g.\ ref.~\cite{CruzMoreno:2013}).
The observation level is set to sea level. We consider SIBYLL~2.3d, EPOS-LHC and Pythia~8.307 as high-energy interaction model above \SI{63.1}{\GeV}.
In each case, Pythia is used as low-energy model down to the hadron/muon cut energy that is set to \SI{1}{\GeV}.
For nuclear projectiles that cannot be treated directly in the simplified nuclear model, we employ the semi-superposition
model implemented in SIBYLL that breaks down $A-A$ collisions into multiple $p/n-A$ collisions~\cite{Engel:1992vf}.

\section{Results}
\Cref{fig:longprof} shows the average longitudinal profiles of proton-induced showers. We observe that Pythia
produces some \SI{30}{\percent} more hadrons than the traditional models in its maximum, which occurs
\SIrange{120}{140}{\gram\per\square\cm} earlier than with the other models. Regarding the muon profile,
the maximum with Pythia is more than \SI{200}{\gram\per\square\cm} shallower and the number of muons at the
maximum is $\sim\SI{6}{\percent}$ higher than with QGSJet-II.04. At ground, however, the number of muons $\Nmu$
is smaller with Pythia due to the longer propagated distance causing higher energy losses and a higher probablity
of in-flight decay. The electromagnetic (EM) longitudinal profiles differ less severly as the EM cascade quickly
decouples from the hadronic one, so that deviations in the hadronic interactions cannot accumulate that much.

\begin{figure}[bt]
\centering
\includegraphics[width=.99\columnwidth]{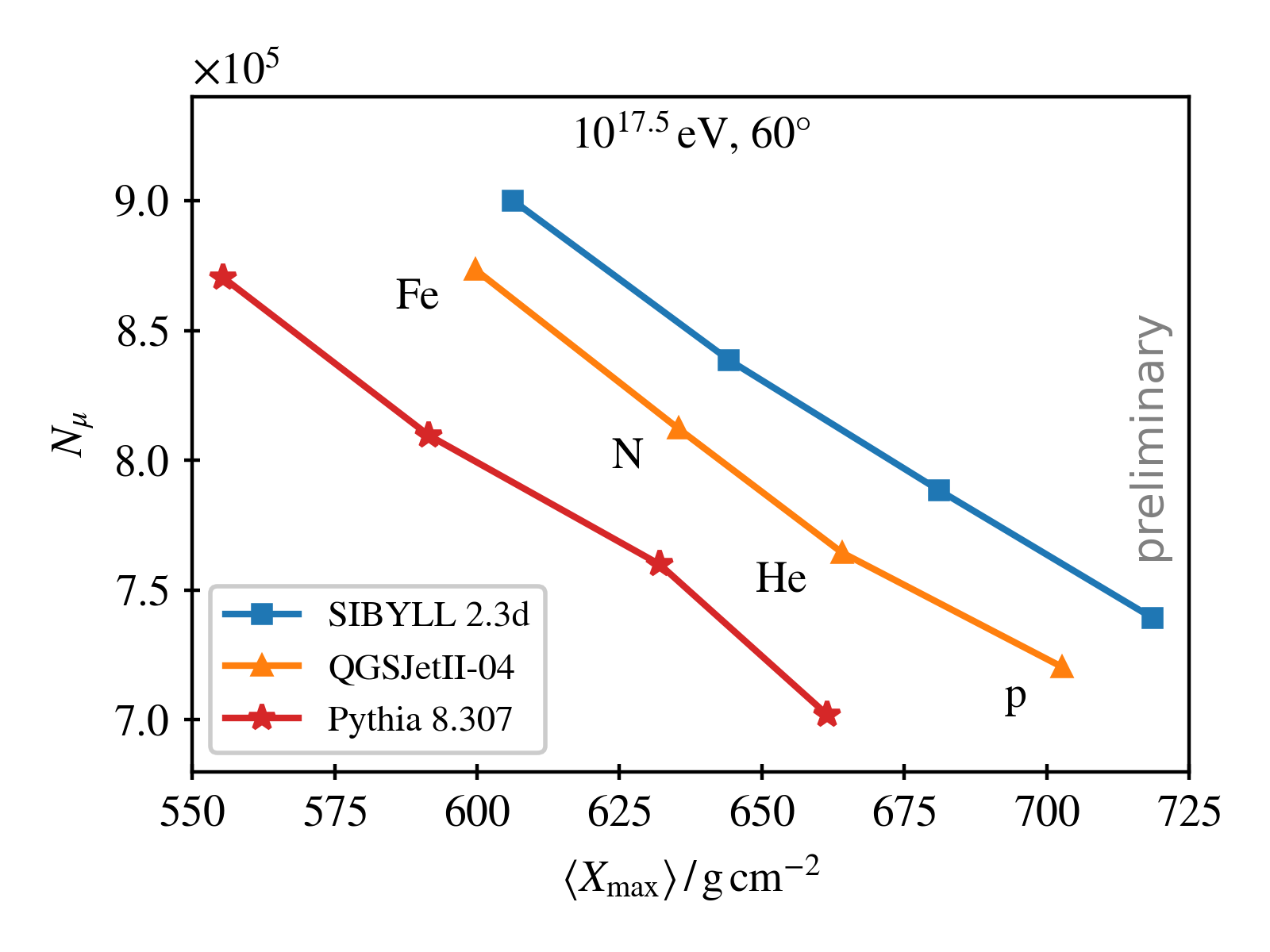}
\caption{Number of muons at ground $\Nmu$ vs.\ shower maximum $\Xmax$}\label{fig:xmax-nmu}
\end{figure}

\Cref{fig:xmax-nmu} displays the results in the $\Xmax$-$\Nmu$ plane also for heavier primaries. We note a significant
shift of the Pythia line by $\sim$\SIrange{40}{50}{\gram\per\square\cm} towards lower $\Xmax$ values w.r.t.\ the other models.

\begin{figure*}[tb]
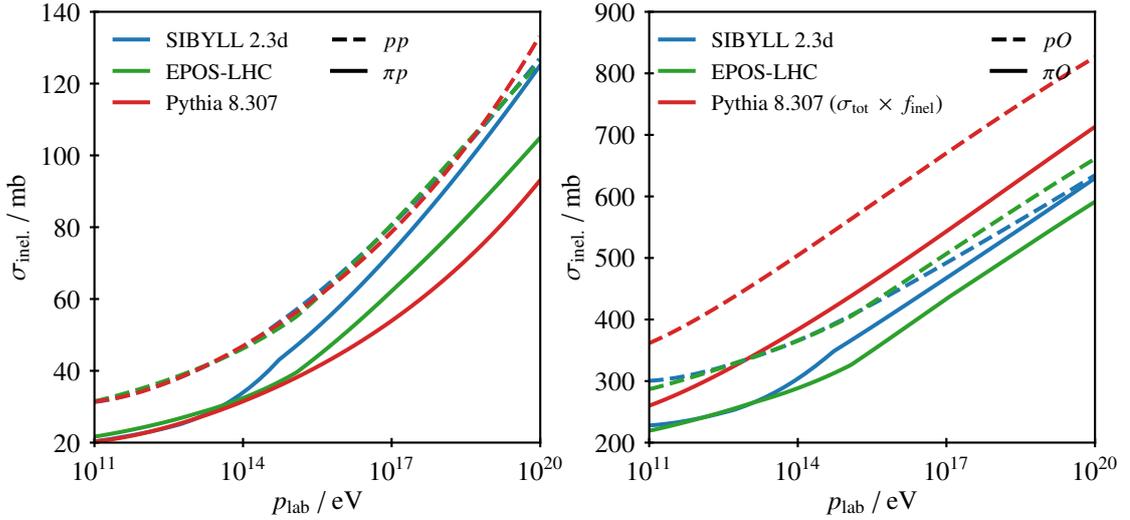

\centering
\begin{minipage}[b]{.49\textwidth}\centering
\input{figures/xs_inel_p.pgf}
\end{minipage}
\begin{minipage}[b]{.49\textwidth}\centering
\input{figures/xs_inel_O.pgf}
\end{minipage}
\caption{Inelastic cross-section predictions. Left: proton target, right: Oxygen target}
\label{fig:cross-sections}
\end{figure*}

It has been shown by Ulrich et al.~\cite{Ulrich:2010rg} that a variation of the inelastic hadron-air cross-sections
has a large impact on $\Xmax$ while leaving $\Nmu$ almost unchanged. For that reason, we show the model predictions
of the inelastic cross-sections in \cref{fig:cross-sections}. Due to the wealth of precise data on $pp$ collisions from
accelerator measurements up to LHC energies ($\sim 10^{17}\,\si{\eV}$ in lab frame) to which the models have been tuned,
the predictions differ only slightly. Precise data on $\pi^\pm p$ collisions, however, exist only up a few \SI{100}{\GeV},
so that predictions diverge especially above $10^{14}\,\si{\eV}$, with Pythia yielding the lowest values. At present, it is unknown whether the $\pi p$ cross-sections eventually converge to the $pp$ ones, which is expected assuming a universal saturation of low-$x$ gluons, or stay below as expected from a Pomeron-style rise.

When considering
oxygen targets, the picture is different. Pythia predicts cross-sections significantly higher than the other models.
The simplified nuclear model of ref.~\cite{Sjostrand:2019zhc} considers only total cross-sections $\sigma_{\mathrm{tot}}$ by employing
the relation $\sigma_{\mathrm{tot}}^{(hA)} = A \sigma_{\mathrm{tot}}^{(hp)} / \langle n_{\mathrm{subcoll}} \rangle$, with
the mean number of subcollisions $\langle n_{\mathrm{subcoll}} \rangle$ parameterized from full Angantyr events. Therefore,
we estimate the inelastic cross-section by scaling $\sigma_{\mathrm{tot}}$ with the ratio $f_{\mathrm{inel}}$ of inelastic
events, which we determined to be approx.\ \SI{92}{\percent} in case of $\pi O$ and \SI{90}{\percent} in case of $p O$ events
with negligible energy dependence. It is noteworthy that Pythia yields the smallest cross-sections among the considered models
in case of $\pi p$ but the largest in case of $\pi O$ and $p O$.

\section{Conclusions and outlook}
We have integrated the latest version of Pythia~8 into CORSIKA~8 to be used as hadronic interaction model
for realistic air shower simulations for the first time. The results presented demonstrate that Pythia is capable to meet
the higher demands (more projectile/target configurations, extrapolations to beyond-LHC energies)
of such simulations compared to its original use-case in accelerator-based high-energy physics.
The observed differences in the longitudinal development can be attributed to cross-section predictions significantly higher than those of the other models -- an issue that requires further investigation and
improved modelling. Further refinements and improvements, also regarding the use of Angantyr directly, are ongoing and expected in upcoming releases.

The availability of Pythia~8 in air shower simulations does not only provide yet another interaction model
but also interesting opportunities: The possibility of tuning the model by the users themselves may offer
new insights into the production of muons and its uncertainties by systematically studying the impact
on air shower observables and accelerator measurements at the same time. Moreover, Pythia~8 is the only
model treating the production of all quark flavors. Until now, only SIBYLL models charm production.
Finally, the advent of Pythia~8, being an object-oriented C++ code, marks an important step towards
enabling parallelization of CORSIKA~8 simulations by multithreading.

\section*{Acknowledgements}
The authors acknowledge support by the High Performance and Cloud Computing Group at the
Zentrum für Datenverarbeitung of the University of Tübingen, the state of Baden-Württemberg through bwHPC
and the German Research Foundation (DFG) through grant no.\ INST~37/935-1~FUGG.

\printbibliography

\end{document}